%% file: VideoAnnotationTool.tex
\documentclass[conference]{IEEEtran}
\IEEEoverridecommandlockouts
\usepackage{cite}
\usepackage{amsmath,amssymb,amsfonts}
\usepackage{algorithmic}
\usepackage{graphicx}
\usepackage{textcomp}
\usepackage{xcolor}
\usepackage{import}
\usepackage{float}
\usepackage{subcaption}
\def\BibTeX{{\rm B\kern-.05em{\sc i\kern-.025em b}\kern-.08em
    T\kern-.1667em\lower.7ex\hbox{E}\kern-.125emX}}

\usepackage{color}    
\newcommand{\comment}[1]{}
\usepackage{hyperref} 

\usepackage[shortlabels]{enumitem}

\usepackage[affil-it]{authblk}
\makeatletter
\renewcommand\AB@affilsepx{, \protect\Affilfont}
\makeatother

\usepackage[font=small,labelfont=bf]{caption}
\begin{document}

\title{An Interactive Annotation Tool\\ for Perceptual Video Compression}

\author[1,*]{Evgenya Pergament}
\author[1]{Pulkit Tandon}
\author[2]{Kedar Tatwawadi}
\author[2]{Oren Rippel}
\author[2]{Lubomir Bourdev}
\author[3]{\\Bruno Olshausen}
\author[1]{Tsachy Weissman}
\author[1]{Sachin Katti}
\author[2]{Alexander G. Anderson}

\affil[1]{Stanford University}
\affil[2]{WaveOne, Inc.}
\affil[3]{UC Berkeley}
\affil[*]{Corresponding author: evgenyap@stanford.edu}

\maketitle

\begin{abstract}

Human perception is at the core of lossy video compression and yet, it is challenging to collect data that is sufficiently dense to drive compression. In perceptual quality assessment, human feedback is typically collected as a single scalar quality score indicating preference of one distorted video over another. In reality, some videos may be better in some parts but not in others. We propose an approach to collecting finer-grained feedback by asking users to use an interactive tool to directly optimize for perceptual quality given a fixed bitrate. To this end, we built a novel web-tool which allows users to paint these spatio-temporal importance maps over videos. The tool allows for interactive successive refinement: we iteratively re-encode the original video according to the painted importance maps, while maintaining the same bitrate, thus allowing the user to visually see the trade-off of assigning higher importance to one spatio-temporal part of the video at the cost of others. We use this tool to collect data in-the-wild (10 videos, 17 users) and utilize the obtained importance maps in the context of x264 coding to demonstrate that the tool can indeed be used to generate videos which, at the same bitrate, look perceptually better through a subjective study --- and are 1.9 times more likely to be preferred by viewers. The code for the tool and dataset can be found at \url{https://github.com/jenyap/video-annotation-tool.git}
\end{abstract}


\section{Introduction}
\import{sections/}{1_Introduction.tex}

\section{Related Work}
\import{sections/}{2_related_work.tex}

\section{Interactive Spatio-Temporal Annotation Tool}\label{sec:annotation_tool}
\import{sections/}{3_annotation_tool.tex}

\section{Experiments and Results}
\import{sections/}{4_results.tex}

\section{Discussion and Future Work}
\import{sections/}{5_discussion.tex}

\section{Acknowledgements}
We would like to thank Ayal Mittelman and Itamar Raviv for helping to develop the annotation tool. 

\bibliographystyle{IEEEtran}
\bibliography{citations.bib}

\end{document}

%% file: sections/1_Introduction.tex
\begin{figure*}
\centering
\begin{subfigure}[t]{\columnwidth}
\includegraphics[width=\columnwidth]{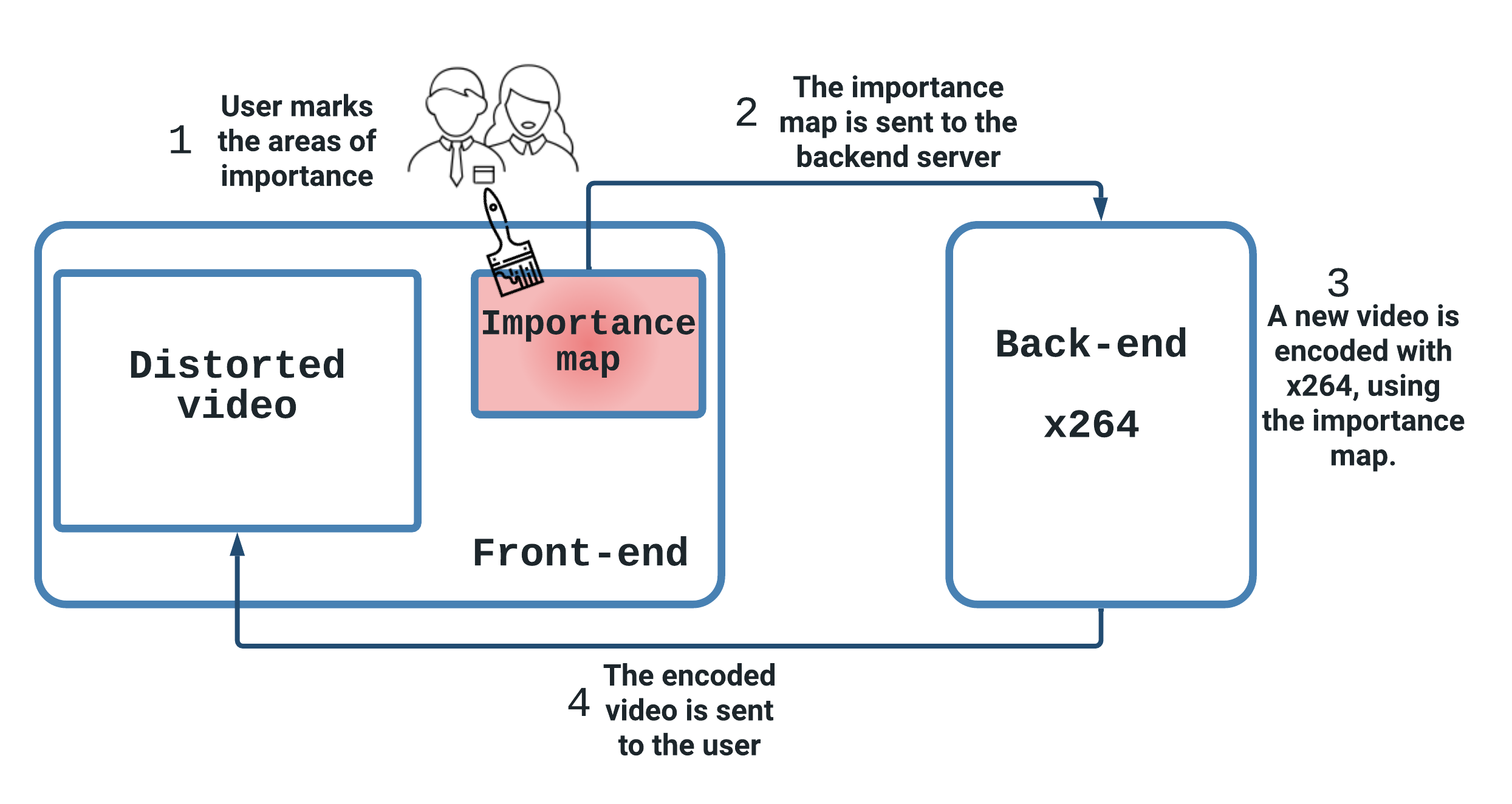}
\caption{Annotation Tool Block Diagram}
\label{fig1:diagram}
\end{subfigure}
\hfill
\begin{subfigure}[t]{\columnwidth}
\includegraphics[width=\columnwidth]{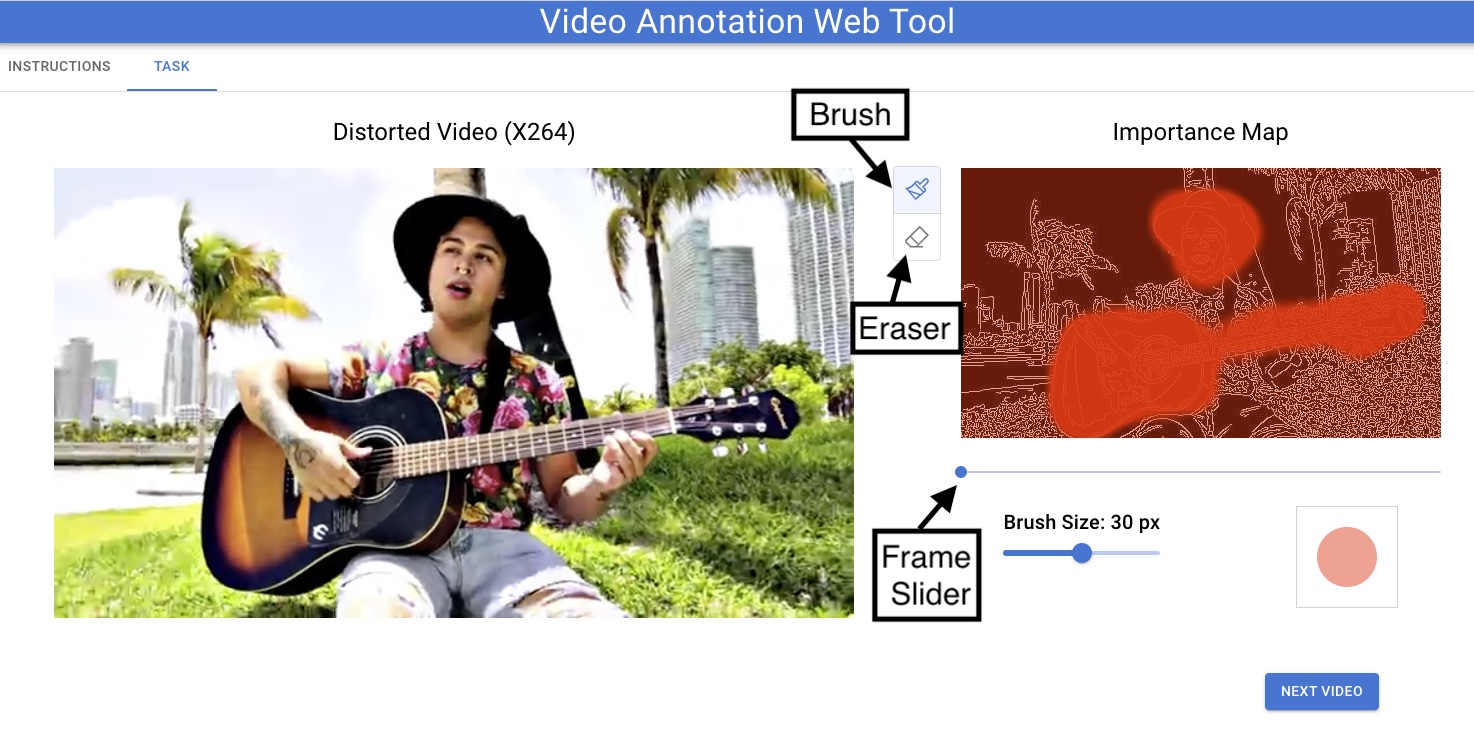}
\caption{Front-end Design}
\label{fig1:frontEnd}
\end{subfigure}\hfill
\caption{An overview of our interactive spatio-temporal importance annotation tool. The tool allows for iterative refinement of the perceptual quality of a video at a fixed bitrate budget, by allowing the user to mark areas of importance through painting an importance map. {\bf (a)} At the front-end, the user can see both the distorted video and current importance map. When the user paints over the importance map [1], the updated importance map is sent to the back-end server [2], where the video is re-encoded with x264 based on the received importance map at the pre-specified bitrate [3]. The updated video is sent back to the front-end [4], allowing the user to immediately see the effect of the painting step and visualize the trade-off of assigning more bits to one region over another. {\bf (b)} The front-end has tools for painting and erasing the importance maps. The painted importance maps are propagated to future frames using optical flow. Users can jump to a particular frame using the frame slider, and pick an appropriate brush size for painting over the video. See Section \ref{sec:annotation_tool} for more detail.\vspace{-0.2in}}
\label{fig1:tool}
\end{figure*}

More than 80\% of the internet comprises video, and this fraction is growing by the day\footnote{Cisco, ”Cisco visual networking index: global mobile data traffic forecast update, 2017–2022.”, \url{https://s3.amazonaws.com/media.mediapost.com/uploads/CiscoForecast.pdf}}. The landscape of video has been evolving quickly over the last few years, spanning a multitude of use-cases such as VOD streaming, social networking, live communication, and remote learning. While the applications are diverse, many feature an important common denominator: the videos are ultimately consumed by humans. This has motivated decades of work on perceptual quality assessment, perceptual quality metrics, and perceptual compression (see \cite{chen2010perceptual, lee2012perceptual} for reviews). However, in order to generate videos better suited for human consumption, a key question remains: what is the best way to collect data in order to perform better perceptual video compression? 

The fields of image and video quality assessment have refined the art and science of collecting subjective data via human feedback. However, the output is usually only a scalar value per comparison. While this certainly provides useful information, it doesn't provide information that is sufficiently specific to drive a video compression algorithm. In order to get spatio-temporally dense importance maps, a related field of work utilizes pre-existing saliency regressors driven by human attention to predict the importance maps which can be used for region-of-interest (ROI) compression \cite{borji2019salient, cong2018review, borji2015salient}. However, this approach only indirectly gets at the core issue, since saliency maps change when a video is compressed.

We introduce a novel approach %
where people use an interactive tool to directly optimize the perceptual quality of a video given a fixed bitrate. The tool allows for collecting finer-grained feedback in the form of spatio-temporal importance maps. We demonstrate using a subjective survey that videos compressed with the human-specified importance maps are 1.9 times more likely to be preferred over videos of the same size compressed in the traditional way. %
We believe this result is just the beginning of this line of research. The data collected through this tool directly tells us which spatio-temporal regions impact the perceptual quality of a video as determined by humans. The importance maps can be used to design new perceptual quality metrics, as well as new compression algorithms based on predicting importance maps.

%% file: sections/2_related_work.tex
Perceptual Quality (PQ) is inherently subjective and as such requires human feedback. The objective of incorporating PQ into the design of video compression and evaluation metrics is a long-standing one, and as such has given rise to a considerable body of literature over built over the span of more than two decades. However, to our knowledge, none of the existing works directly tackle the question of: which spatio-temporal regions do viewers find important in compressed videos? Below we delineate the various related works, which we compare and contrast to ours.

\subsection{Scalar Quality Score Collection}
Several procedures have been developed by the International Telecommunication Union (ITU) to assess the quality of images and videos \cite{brunnstrom2009vqeg, winkler2009video}. The procedures can be roughly split into two main categories: \emph{rating} in absolute terms, or \emph{ranking} in relative terms. In rating, the observer is presented with a video and their task is to rate it with a score. For example, in ACR (Absolute Category Rating) the observer rates a video on a scale from 1 to 5 to reflect asset quality. In ranking, the observer is presented with two or more videos, and they are asked to rank the videos in order of preference. In both of these methods, the output score is only a single scalar, and as such cannot not provide transparency into the critical question as to \emph{why} each user provided this response. For instance, was it due to noticeable distortions to specific areas of interest? In our work, we focus on collecting spatio-temporal feedback on what an individual finds important in the video.

\begin{figure*}
\begin{center}
\includegraphics[width=0.8\textwidth]{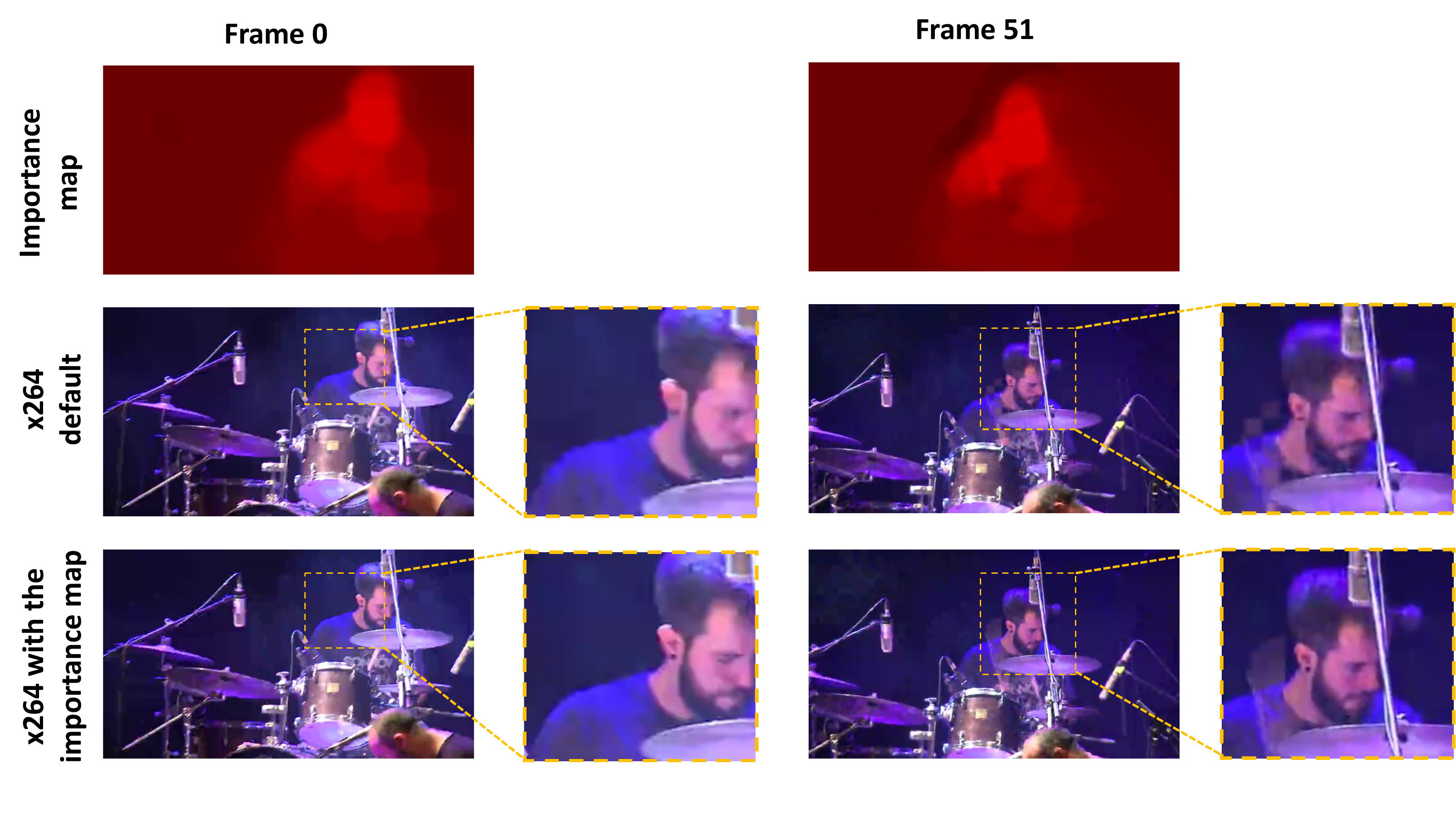}
\vspace{-0.2in}
\caption{Example of user-annotated importance map frames, frames from video encoded with the default x264 implementation, and frames from video encoded with x264 using the importance maps. In this case, it can be seen that the reconstructions are sharper and have higher quality in the areas that were assigned with high importance. See Sections \ref{sec:importance_map} and \ref{sec:x264_compression} for more detail.\vspace{-0.2in}}
\label{fig:frames_via_tool}
\end{center}
\end{figure*}

\subsection{Saliency Map Collection}
Predicting visual attention has been a field with long-standing interest, due to its widespread applications to many fields such as image/video quality assessment \cite{image-quality, gu2016saliency}, image/video compression \cite{han2006image, saliency-preserving-video-compression}, and so on. In the past three decades, numerous works explored this question from different directions (see \cite{borji2019salient, cong2018review, borji2015salient} for reviews). As humans can find different regions visually salient in a scene given the context, these approaches can be classified as either bottom-up \cite{Itti-saliency-based, cheng2014global, li2014secrets, lei2016universal, shi2015hierarchical} or top-down \cite{liu2010learning, zhang2018deep, zhang2017learning, wang2017learning}. In top-down saliency detection, human visual saliency is detected under a given high-level task, whereas bottom-up saliency detection is stimulus-driven with the goal of identifying important visual features under more natural settings. In our work, we take a top-down approach, since lossy video compression introduces characteristic but undesirable artifacts, such as blocking, ringing and aliasing, and these artifacts can potentially shift the observer's attention from the original regions of interest during iterative painting. 

On the other hand, based on how the saliency maps are collected, the majority of the saliency literature focuses on either fixation prediction \cite{judd2009learning, wilming2017extensive, borji2015cat2000, fan2018emotional} or salient object detection \cite{cheng2014global, liu2010learning, jiang2011automatic, hou2017deeply}. In fixation experiments, saliency is measured by tracking human gaze whereas in salient object detection, the goal is to find pixel-accurate segment maps of important objects in a scene. Unfortunately, gaze tracking is inherently non-scalable as it requires special equipment in a laboratory setting and most of the salient object detection focus on detection of one or few salient objects instead of the whole scene. In our approach, we allow unconstrained painting within the entire spatio-temporal volume by the user, enabling us to a collect a fine-grained data at pixel level while being uninhibited by the limited segmented regions in the video.

Finally, many approaches have tried to exploit one of the many saliency models (or built their own) for the specific task of video compression \cite{saliency-Aware-video-compression, saliency-preserving-video-compression, xu2014region, gupta2013visual, pvc-stim}. However, none of these methods directly collect the human spatio-temporal saliency maps explicitly for the task of video compression. In our work, we collect perceptual data from distorted videos in an interactive successive refinement process where the user provides immediate feedback on how the chosen saliency region impacts the encoded video --- which, importantly, maintains any resultant compression artifacts within consideration. 

%% file: sections/3_annotation_tool.tex
The annotation tool allows a user to iteratively refine a spatio-temporal importance map in order to improve the perceptual quality of an H.264-encoded video given a fixed bitrate budget. Below we present the workflow of the tool, the considerations in its design, and the implementation details. %

\subsection{Annotation Tool Workflow} 

Fig. \ref{fig1:diagram} shows the workflow of the tool. The user is presented with the H.264-encoded video on the left, and an annotation map on the right. For simplicity we will refer to the frames on the right as annotation map frames. Each such frame could be thought of as a canvas where the user marks particular areas where they want higher quality (i.e. lower distortion). The annotation tool allows the user to paint areas of importance in the video in an iterative fashion. Each iteration breaks down into 4 steps:
\begin{enumerate}[{\bf S1:}]
    \item In the first step, the user paints areas of the annotation map frames where they want to see better quality (or conversely erase areas where higher quality is not needed). We refer to the result of the user's marking as an \emph{importance map} (Section \ref{sec:importance_map}). 
    \item The user-generated importance map is sent over to the back-end server, to re-generate the encoded video based on the importance map. 
    \item The backend-server re-generates the encoded video based on the received importance map using the x264 video encoder, at a fixed bitrate (Section \ref{sec:x264_compression}). 
    \item When the encoding is finished, the encoded video is sent back to the browser. On the left, the user is presented with the newly re-encoded video, and on the right with the updated importance map (which is temporally propagated to facilitate future frame annotation; see Section \ref{sec:temporal_propagation}). This allows the user immediate feedback on the annotation and allows for iterative refinement of the importance map. 
\end{enumerate}

The user repeats steps S1-S4 until they are content with the result. Once the user presses on the `Next Video' button (Fig. \ref{fig1:frontEnd}), the final importance map is saved on the back-end and a new video is presented. 

\subsection{Importance Map Specification and Annotation}\label{sec:importance_map}
The importance map is available for the user to refine for each frame, and its values range between 0 to 255, where 0 is represented with black and 255 with red. The redder an area is, the more important it is to the user. To start, all pixels of the annotation map are initialized to 127. We also overlay source video edges on the annotation map to assist with the painting. 

The user can either increase the importance of an area by using the brush, or decrease it by using the eraser (Fig. \ref{fig1:frontEnd}). The brush size can be adjusted, allowing the user to mark large areas quickly, or conversely to focus on finer-grained details. Both the coloring and erasing operations are implemented using additive functions, which spatially decay with the brush radius.
In order to dissuade the user from coloring the entire video in red (255), the importance map is normalized to an average pixel value of 127 and the normalized importance map is presented on the right.

\subsection{Temporal Propagation of the Annotation Map}\label{sec:temporal_propagation}
\begin{figure}
\includegraphics[width=\columnwidth]{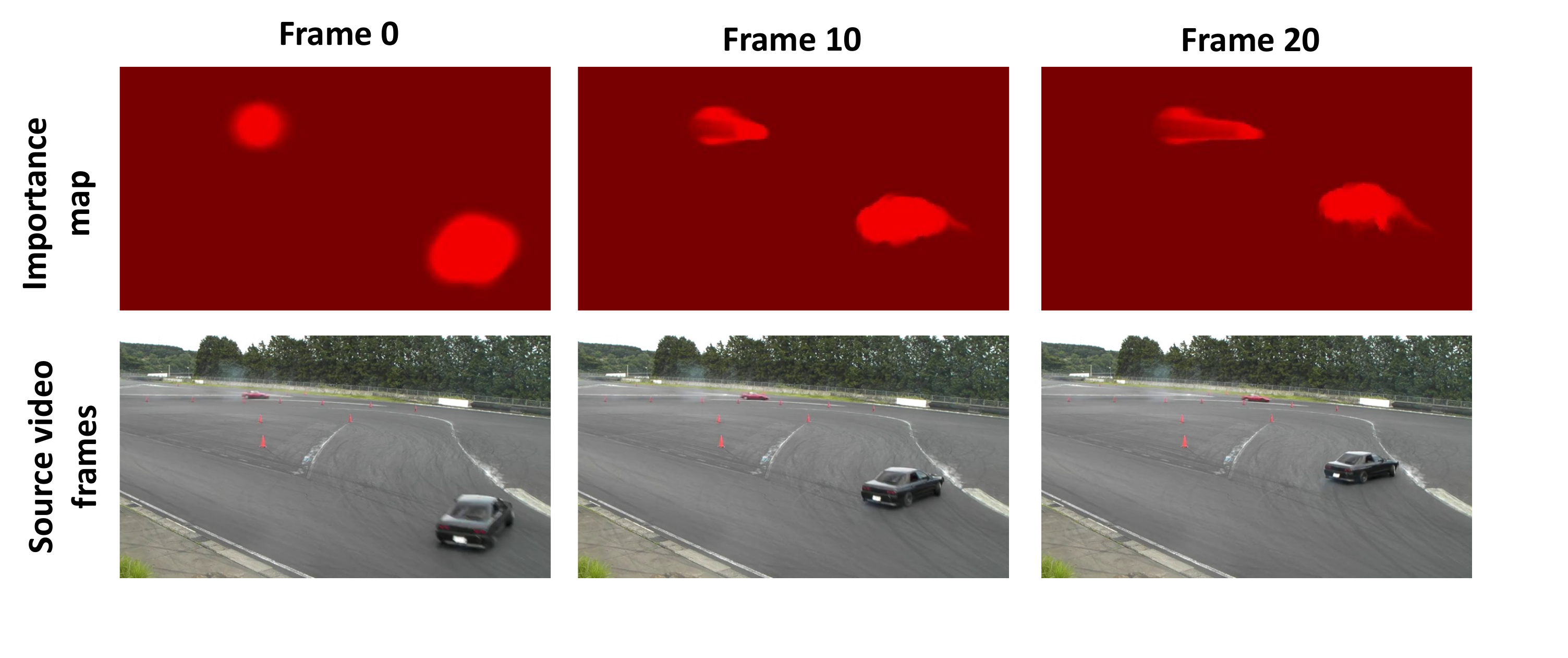}
\caption{After the user paints a particular importance map, the map is propagated to future frames using optical flow. This facilitates the task of painting the mask, as otherwise the user would have to paint the mask from scratch in every frame (see Section \ref{sec:temporal_propagation}).\vspace{-0.2in}}
\label{fig2:flow_propagation}
\end{figure}

To facilitate the annotation, each time the user paints on the canvas, we propagate their marking to future frames. If a user marks a certain area/object as important, we expect it to remain important for the next few frames. To that end, we estimate the optical flow between consecutive frames, and utilize this estimate to propagate the paintings between frames. An example of such propagation is shown in Fig. \ref{fig2:flow_propagation}.

As the optical flow estimation algorithms are not perfect and neither is the user-generated mask, this can sometimes lead to errors in the annotation map propagation. Note, however, that the user can manually correct the annotation map in such a scenario. To mitigate the errors in the optical flow estimation, the annotation map is only propagated to the next 40 of frames, and is temporally decayed. For efficiency, the optical flow is pre-computed and stored on the back-end server. We used the off-the-shelf OpenCV implementation of the Farneback algorithm to estimate the optical flow\footnote{OpenCV Farneback implementation: \url{https://docs.opencv.org/3.4/d4/dee/tutorial_optical_flow.html}}.

\subsection{H.264/AVC Compression Standard and Implementation}\label{sec:x264_compression}
In order to optimize perceptual quality given a fixed budget, videos are compressed into H.264-compliant bitstreams \cite{H264-Overview, H264-standard, H264-from-concepts} using x264 (the most commonly used H.264 encoder)\footnote{L. Merritt and R. Vanam, “x264: A high performance h.264/avc
encoder”, \url{https://citeseerx.ist.psu.edu/viewdoc/summary?doi=10.1.1.96.8042}}.  The relative effort (i.e. bits) spent on different parts of the frames of the video is modulated by the $QP$ parameter in the internals of x264. This parameter in turn impacts the mode selection and the residual coding of each macroblock. We have modified the x264 code to receive as input a $\Delta QP$-map, which indicates how to modify the $QP$ value of each macroblock. The $\Delta QP$-map is a linear function of the importance map, mapping normalized importance map between [0, 1] to a $\Delta QP$ of [-10, 10]. 

To elaborate, H.264 divides each frame of the video into 16x16 macroblocks. There are different modes for coding each macroblock, such as using motion vectors to generate an initial prediction of the block, predicting based on previously-decoded neighbors, and optionally coding a residual. At the start of the analysis of a macroblock, the overall effort spent on that block is modulated by a block-specific \underline{q}uantization \underline{p}arameter (QP) in the x264 encoder. This QP is converted to $\lambda_1$ for mode selection and $\lambda_2$ for residual coding. The mode selection proceeds by evaluating different prediction modes to ascertain which minimizes the rate distortion loss $D_{SATD} + \lambda_1 \cdot R$ where $D_{SATD}$ is the sum of absolute Hadamard transformed differences. Next, the encoder searches over the QP parameter of the residual transform (this is correlated to but not exactly the same as the QP parameter to choose the $\lambda$s) to optimize the $D_{SSD} + \lambda_2 \cdot R$ loss where $D_{SSD}$ is the sum of squared differences loss. The 2-pass ABR (\underline{a}daptive \underline{b}it\underline{r}ate) mode is used with this modification (applied to both passes) to compress the video to a target bitrate. The 2-pass algorithm computes $\Delta QPs$ on top of our externally-specified ones in order to improve compression using the analysis of future frames and to achieve a given target bitrate. The baseline is a 2-pass ABR encoding with a delta QP map of zero, which reduces exactly to a normal 2-pass encoding using x264.

%% file: sections/4_results.tex
\comment{
    \begin{enumerate}
    \color{red}{
    \item Dataset -- videos used
    \item Annotation Map collection
    \item One big figure for all these: for the collected dataset?
    \item Subjective Study 
    \item Figure/Table for subjective study
    }
    \end{enumerate}
}

The dataset, obtained importance map and videos used for the subjective study can be found in our GitHub repository.

\subsection{Video Dataset}
In our evaluations, we used 10 videos, each of 3-5 seconds. The videos were chosen from the Video Compression Challenge in CLIC 2021\footnote{\url{https://storage.googleapis.com/clic2021_public/txt_files/video_urls.txt}}. We chose videos from the categories `Animation', `Gaming', `Lecture', `Live Music', `Sports' and `Vlog'. Each video was resized to match our annotation tool's video dimensions ($800\times450$) and cropped to be less than 5 seconds. All the videos were encoded at a bitrate such that the resulting output has a PSNR of 25.

\subsection{Importance Map Collection}
\label{sec_results_imp_map}
To collect the annotation data, we installed our annotation tool on a Google Cloud Platform (GCP) VM. %
We used Amazon Mechanical Turk (MTurk) to collect the importance map data. MTurk workers were required to have the ``masters'' qualification and lifetime approval rate greater than 98\% to be qualified for our study. Each annotator was asked to annotate 5 random videos from the dataset, and was requested to spend at least 3 minutes on each video. This was enforced by disabling the ``Next video'' button for 3 minutes. The annotation tool provides a unique ID for each user that uses the tool. Using this ID we aggregated all the saved importance map by user. We also collected, for each user, intermediate importance maps (maps recorded while the user was using the tool) and x.264 encoding statics. We disqualified data collected from users who did not complete annotation task --- for example, users who just waited for the `Next video' button to enable or annotated only a small part in just one frame. 

Figure \ref{fig:frames_via_tool} shows example importance maps collected using the tool. It can be seen in the obtained frames that the tool indeed produces higher quality encodes in the regions identified as important by the annotator.

\begin{figure}
\includegraphics[width=\columnwidth]{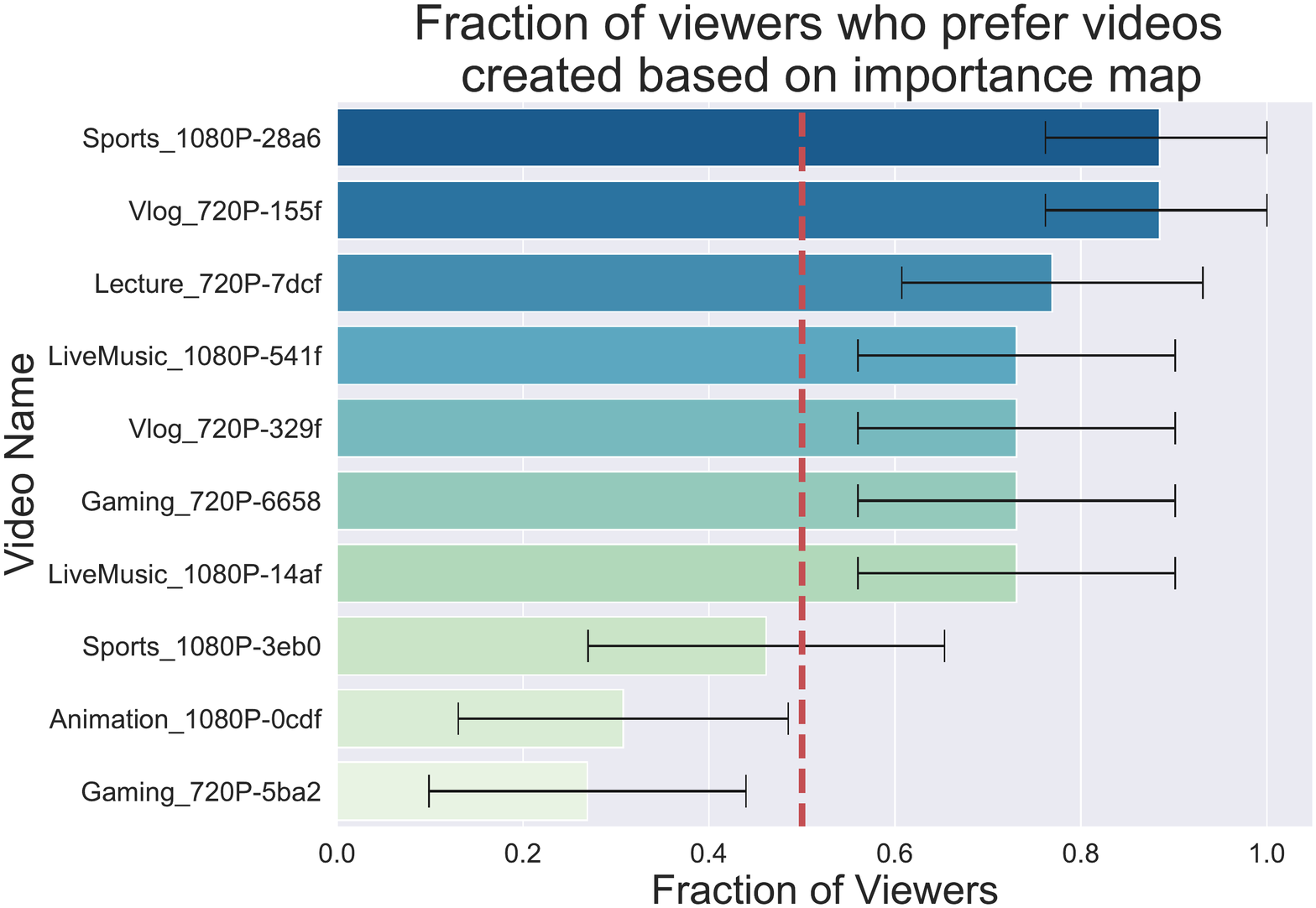}
\caption{Subjective study results. The bar chart shows the fraction of viewers which preferred the video encoded with the human-generated average importance map over the baseline encoding with the same bitrate. The error bars show 95\% confidence interval. The importance-weighted encodes generated by the tool were 1.9 times more likely to be preferred over the baseline x264 encodes across the dataset. See Section \ref{sec:subjective_study} for more detail.\vspace{-0.25in}}
\label{fig:result}
\end{figure}

\subsection{Validation via a Subjective Study}\label{sec:subjective_study}
We validated that the tool functions as desired via a subjective study of the collected importance maps. For each video, we compute an average importance map across annotators. The average importance map is used to create a final video, which is then compared against the baseline encoded video without any spatio-temporal importance at similar bitrate. The encoded baseline and importance-map-generated videos are compared for all examples in the dataset in a separate subjective study involving different subjects ($n=40$) using a two-alternative forced choice (2AFC) approach. The subjective study was generated using the Qualtrics Platform and conducted using Amazon MTurk, with similar requirements on workers as described in Section \ref{sec_results_imp_map}. We also included sanity checks in our comparison: 3 pairs of videos were added to the study such that one amongst the pair was obviously better. Subjects who incorrectly selected even one of these videos with worse quality were rejected from the data analysis. This resulted in $n=26$ subjects whose selections were incorporated in the reported results. 

Figure \ref{fig:result} shows the results for the different input videos. The bar chart shows the fraction of viewers who preferred importance map encoded videos over the baseline videos for different contents. The error bars represent the 95\% Confidence Interval using standard normal distribution assuming each choice can be modeled using a binomial distribution. On average, across the dataset, videos generated using the importance maps were 1.9 times more likely to be chosen over the baseline encoded videos. The importance map-based method outperforms the baseline on 7 out of 10 videos, performs at chance on 1 video, and underperforms on 2 videos. 

\begin{figure}[t]
\includegraphics[width=\columnwidth]{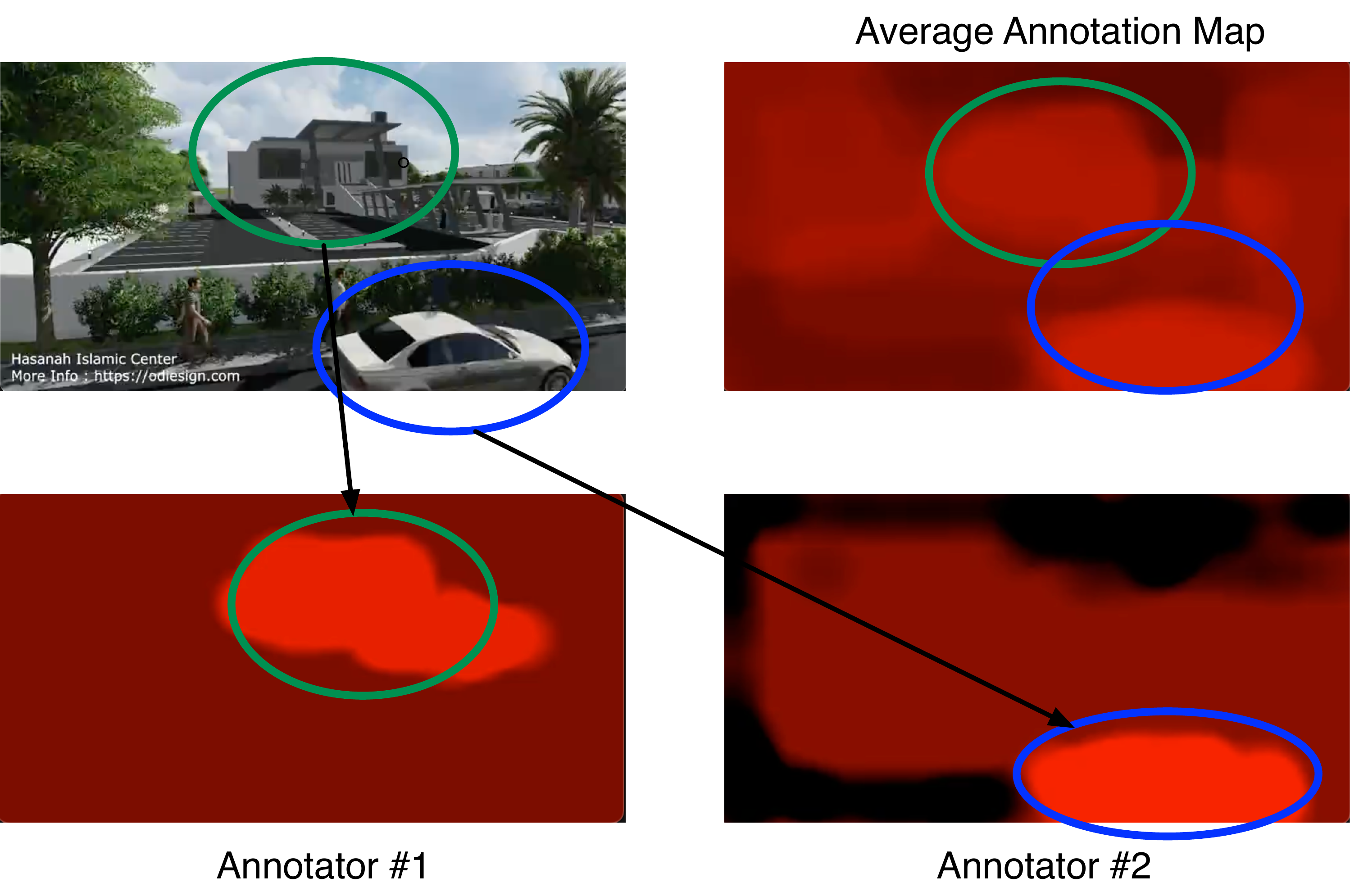}
\caption{Example failure case. For the `Animation 1080P-0cdf' asset, different annotators preferred different objects in the scene (blue and green ovals), resulting in an average map mostly agnostic to both the salient objects. See Section \ref{sec:subjective_study} for an extended discussion.\vspace{-0.2in}}
\label{fig:failure_case_1}
\end{figure}

To understand these failure cases of the tool, we dug deeper into the two videos (`Animation\_1080P-0cdf', `Gaming\_720P-5ba2') where baseline encodes were preferred over the importance map encodes. For the case of `Animation' video, we found that there were two salient objects in the video and different annotators preferred different objects during the importance map collection step, resulting in an average map which was largely agnostic to both the salient maps (Figure~\ref{fig:failure_case_1}). This resulted in a video where neither of the salient objects were preferred and highlights the issue that the average importance maps might lead to a competition between the preferences of different annotators. On the other hand, for the `Gaming' video, we found that the annotators all agreed to focus on the central character which is moving around in an artificial background. However, due to the over-focus on the central character, the video ends up looking unnatural due to the sudden reduction in the quality of the background in the neighborhood surrounding the character. The large motion exacerbates this issue, making it feel like the high quality character is moving in a low quality background. This example highlights the issue that during the annotation task, an annotator might have higher focus on the particular region they are painting and not take into account the increased distortions in unmarked areas, as there is no reference video for them to compare their final encode against.

%% file: sections/5_discussion.tex
In this work, we proposed a tool which allows collecting importance maps directly for the task of perceptual video coding. Furthermore, we demonstrated that re-encoding using these annotations produces videos that are preferred relative to the counterpart baseline encoding. 
This initial success suggests a wide variety of follow-up work that improves the data collection procedure and utilizes the obtained data.

First, the task can be made easier to understand and more can be done to automatically validate the data. For instance, users tend to begin by annotating the main objects in the scene without attending to emergent artifacts in the background. In one such failure case, users annotated a character in the foreground, resulting in significant artifacts in the background. The annotated video received lower scores when directly compared with the original. While we instructed annotators to look carefully at the background after their first pass of annotating, more validation is likely necessary. In terms of validating the data quality, we had to manually exclude users who, for instance, annotated only the first frame or did not perform any annotations at all. One approach would be to present the user their video side by side with the baseline and verify that they choose the video that they annotated. 

Second, it would be informative to decompose the importance maps into various sources of viewer attention --- originating due to the saliency of source content versus artifact-dependent attention. Independent of the underlying compression algorithm, some aspects of the source video catch the attention of the average viewer more or less readily. These may include increased attention on people/text or reduced sensitivity to fast-moving objects. However, compression-dependent artifacts also catch the eyes of the viewers. It would be important to understand the extent to which an importance map collected at one target bitrate is useful for encoding a video at another target bitrate. It is likely that importance based on compression artifacts would change with bitrates, but importance based on the source content would remain consistent.  In our initial experimentation, we found that applying the importance maps to encoding at 1.5$\times$ the collected bitrate resulted in videos which were preferred 1.4 times compared to 1.9 times with original bitrates.
Another important factor of variation is the extent to which the users annotate the videos differently. Some users may attend to different objects than others and others may be more sensitive to compression artifacts. 

Third, it is highly likely that the importance maps can be predicted to a sufficient degree to enable automatic perceptual compression. With a neural network trained on a large dataset of compressed video-importance map pairs, the entire perceptual coding process could be automated. Finally, we believe that the collected importance maps can be useful for the creation of perceptual metrics. Much like \cite{lin2020pea265}, spatial localization of artifacts can be useful for measuring perceptual quality. One way to create a perceptual metric using our data would be to have an importance-map weighted distortion loss. We plan on collecting a large-scale dataset using the tool for automated perceptual compression in future.

Perceptually-driven video compression is a well-known and important problem. This paper explored a novel approach to this problem by asking users to directly optimize the perceptual quality using an interactive tool. 
The crowd-sourced importance maps are leveraged to re-encode the videos and a 2AFC study confirms that users prefer these perceptually-coded videos. This paper lays the foundation for a wide variety of follow up work in the fields of perceptual video coding and perceptual quality assessment.